\documentclass[10pt,a4paper,onecolumn]{article}
\usepackage{marginnote}
\usepackage{graphicx}
\usepackage{xcolor}
\usepackage{authblk,etoolbox}
\usepackage{titlesec}
\usepackage{calc}
\usepackage{tikz}
\usepackage{hyperref}
\hypersetup{colorlinks,breaklinks=true,
            urlcolor=[rgb]{0.0, 0.5, 1.0},
            linkcolor=[rgb]{0.0, 0.5, 1.0}}
\usepackage{caption}
\usepackage{tcolorbox}
\usepackage{amssymb,amsmath}
\usepackage{ifxetex,ifluatex}
\usepackage{seqsplit}
\usepackage{xstring}

\usepackage{float}
\let\origfigure\figure
\let\endorigfigure\endfigure
\renewenvironment{figure}[1][2] {
    \expandafter\origfigure\expandafter[H]
} {
    \endorigfigure
}

\usepackage{fixltx2e} 
\usepackage[
  backend=biber,
]{biblatex}
\bibliography{paper.bib}

\let\textttOrig=\texttt
\def\texttt#1{\expandafter\textttOrig{\seqsplit{#1}}}
\renewcommand{\seqinsert}{\ifmmode
  \allowbreak
  \else\penalty6000\hspace{0pt plus 0.02em}\fi}

\makeatletter
\let\href@Orig=\href
\def\href@Urllike#1#2{\href@Orig{#1}{\begingroup
    \def\Url@String{#2}\Url@FormatString
    \endgroup}}
\def\href@Notdoi#1#2{\def\tempa{#1}\def\tempb{#2}%
  \ifx\tempa\tempb\relax\href@Urllike{#1}{#2}\else
  \href@Orig{#1}{#2}\fi}
\def\href#1#2{%
  \IfBeginWith{#1}{https://doi.org}%
  {\href@Urllike{#1}{#2}}{\href@Notdoi{#1}{#2}}}
\makeatother

\newlength{\cslhangindent}
\newlength{\csllabelwidth}
\newenvironment{CSLReferences}[3] 
 {
  \setlength{\parindent}{0pt}
  \ifodd #1 \everypar{\setlength{\hangindent}{\cslhangindent}}\ignorespaces\fi
  \ifnum #2 > 0
  \setlength{\parskip}{#2\baselineskip}
  \fi
 }%
 {}
\usepackage{calc}

\usepackage[top=3.5cm, bottom=3cm, right=1.5cm, left=1.0cm,
            headheight=2.2cm, reversemp, includemp, marginparwidth=4.5cm]{geometry}

\titleformat{\section}
  {\normalfont\sffamily\Large\bfseries}
  {}{0pt}{}
\titleformat{\subsection}
  {\normalfont\sffamily\large\bfseries}
  {}{0pt}{}
\titleformat{\subsubsection}
  {\normalfont\sffamily\bfseries}
  {}{0pt}{}
\titleformat*{\paragraph}
  {\sffamily\normalsize}

\usepackage{fancyhdr}
\makeatletter
\let\ps@plain\ps@fancy
\fancyheadoffset[L]{4.5cm}
\fancyfootoffset[L]{4.5cm}

\definecolor{linky}{rgb}{0.0, 0.5, 1.0}

\newtcolorbox{repobox}
   {colback=red, colframe=red!75!black,
     boxrule=0.5pt, arc=2pt, left=6pt, right=6pt, top=3pt, bottom=3pt}

\patchcmd{\@maketitle}{center}{flushleft}{}{}
\patchcmd{\@maketitle}{center}{flushleft}{}{}
\patchcmd{\@maketitle}{\LARGE}{\LARGE\sffamily}{}{}
\def\maketitle{{%
  
  \AB@maketitle}}
\makeatletter
\renewcommand\AB@affilsepx{ \protect\Affilfont}
\renewcommand\AB@affilnote[1]{{\bfseries #1}\hspace{3pt}}
\renewcommand{\affil}[2][]%
   {\newaffiltrue\let\AB@blk@and\AB@pand
      \if\relax#1\relax\def\AB@note{\AB@thenote}\else\def\AB@note{#1}%
        \setcounter{Maxaffil}{0}\fi
        \begingroup
        \let\href=\href@Orig
        \let\texttt=\textttOrig
        \let\protect\@unexpandable@protect
        \def\thanks{\protect\thanks}\def\footnote{\protect\footnote}%
        \@temptokena=\expandafter{\AB@authors}%
        {\def\\{\protect\\\protect\Affilfont}\xdef\AB@temp{#2}}%
         \xdef\AB@authors{\the\@temptokena\AB@las\AB@au@str
         \protect\\[\affilsep]\protect\Affilfont\AB@temp}%
         \gdef\AB@las{}\gdef\AB@au@str{}%
        {\def\\{, \ignorespaces}\xdef\AB@temp{#2}}%
        \@temptokena=\expandafter{\AB@affillist}%
        \xdef\AB@affillist{\the\@temptokena \AB@affilsep
          \AB@affilnote{\AB@note}\protect\Affilfont\AB@temp}%
      \endgroup
       \let\AB@affilsep\AB@affilsepx
}
\makeatother

\renewcommand\Affilfont{\sffamily\small\mdseries}
\ifnum 0\ifxetex 1\fi\ifluatex 1\fi=0 
  \usepackage[T1]{fontenc}
  \usepackage[utf8]{inputenc}

\else 
  \ifxetex
    \usepackage{mathspec}
    \usepackage{fontspec}

  \else
    \usepackage{fontspec}
  \fi
  \defaultfontfeatures{Ligatures=TeX,Scale=MatchLowercase}

\fi
\IfFileExists{upquote.sty}{\usepackage{upquote}}{}
\IfFileExists{microtype.sty}{%
\usepackage{microtype}
\UseMicrotypeSet[protrusion]{basicmath} 
}{}

\usepackage{hyperref}
\hypersetup{unicode=true,
            pdftitle={KUIELab-MDX-Net: A Two-Stream Neural Network for Music Demixing},
            pdfborder={0 0 0},
            breaklinks=true}
\usepackage{longtable,booktabs}

\let\addcontentslineOrig=\addcontentsline
\def\addcontentsline#1#2#3{\bgroup
  \let\texttt=\textttOrig\addcontentslineOrig{#1}{#2}{#3}\egroup}
\let\markbothOrig\markboth
\def\markboth#1#2{\bgroup
  \let\texttt=\textttOrig\markbothOrig{#1}{#2}\egroup}
\let\markrightOrig\markright
\def\markright#1{\bgroup
  \let\texttt=\textttOrig\markrightOrig{#1}\egroup}

\usepackage{graphicx,grffile}
\makeatletter
\def\maxwidth{\ifdim\Gin@nat@width>\linewidth\linewidth\else\Gin@nat@width\fi}
\def\maxheight{\ifdim\Gin@nat@height>\textheight\textheight\else\Gin@nat@height\fi}
\makeatother
\setkeys{Gin}{width=\maxwidth,height=\maxheight,keepaspectratio}
\IfFileExists{parskip.sty}{%
\usepackage{parskip}
}{
\setlength{\parindent}{0pt}
\setlength{\parskip}{6pt plus 2pt minus 1pt}
}
\providecommand{\tightlist}{%
  \setlength{\itemsep}{0pt}\setlength{\parskip}{0pt}}
\ifx\paragraph\undefined\else
\let\oldparagraph\paragraph
\renewcommand{\paragraph}[1]{\oldparagraph{#1}\mbox{}}
\fi
\ifx\subparagraph\undefined\else
\let\oldsubparagraph\subparagraph
\renewcommand{\subparagraph}[1]{\oldsubparagraph{#1}\mbox{}}
\fi

\title{KUIELab-MDX-Net: A Two-Stream Neural Network for Music Demixing}

        \author[1]{Minseok Kim\footnote{co-first author}}
          \author[2]{Woosung Choi\footnote{co-first author}}
          \author[3]{Jaehwa Chung}
          \author[4]{Daewon Lee}
          \author[1]{Soonyoung Jung\footnote{corresponding author}}

      \affil[1]{Korea University}
      \affil[2]{Queen Mary University of London}
      \affil[3]{Korea National Open University}
      \affil[4]{Seokyeong University}
  \date{\vspace{-7ex}}

\begin{document}
\maketitle

\marginpar{

  \begin{flushleft}
  \sffamily\small

  \vspace{2mm}

  \par\noindent\hrulefill\par

  \vspace{2mm}

  \vspace{2mm}
  {\bfseries License}\\
  Authors of papers retain copyright and release the work under a Creative Commons Attribution 4.0 International License (\href{http://creativecommons.org/licenses/by/4.0/}{\color{linky}{CC BY 4.0}}).

  \vspace{4mm}
  {\bfseries In partnership with}\\
  \vspace{2mm}
  \includegraphics[width=4cm]{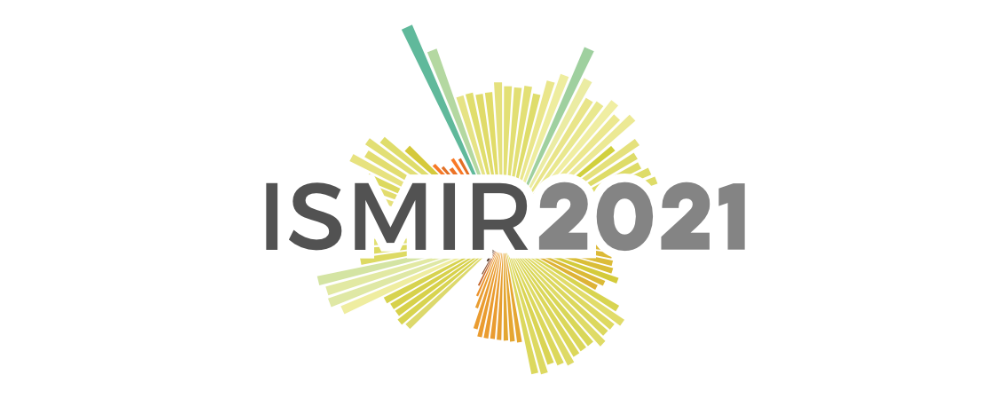}

  \end{flushleft}
}

\hypertarget{summary}{%
\section{Summary}\label{summary}}

Recently, many methods based on deep learning have been proposed for
music source separation. Some state-of-the-art methods have shown that
stacking many layers with many skip connections improve the SDR
performance. Although such a deep and complex architecture shows
outstanding performance, it usually requires numerous computing
resources and time for training and evaluation. This paper proposes a
two-stream neural network for music demixing, called KUIELab-MDX-Net,
which shows a good balance of performance and required resources. The
proposed model has a time-frequency branch and a time-domain branch,
where each branch separates stems, respectively. It blends results from
two streams to generate the final estimation. KUIELab-MDX-Net took
second place on leaderboard A and third place on leaderboard B in the
Music Demixing Challenge at ISMIR 2021. This paper also summarizes
experimental results on another benchmark, MUSDB18.
Our source code is available online \footnote{https://github.com/kuielab/mdx-net}.

\hypertarget{introduction}{%
\section{Introduction}\label{introduction}}

Recently, many methods have been proposed for music source separation.
Notably, deep learning approaches Défossez et al. (2021) have become
mainstream because of their excellent performance. Some state-of-the-art
methods (Choi et al., 2020; Takahashi et al., 2018; Takahashi \&
Mitsufuji, 2021, 2017) have shown that stacking many layers with many
skip connections improve the SDR performance.

Although a deep and complex architecture shows outstanding performance,
it usually requires numerous computing resources and time for training
and evaluation. Such disadvantages make them not affordable in a
restricted environment where limited resources are provided. For
example, some deep models such as LaSAFT-Net (Choi et al., 2021) exceed
the time limit of the Music Demixing Challenge (MDX) at ISMIR 2021
(Mitsufuji et al., 2021) even if they are the current state of the art
on the MUSDB18 (Rafii et al., 2017b) benchmark.

This paper presents a source separation model named KUIELab-MDX-Net. We
empirically found a good balance of performance and required resources
to design KUIElab-MDX-Net. For example, we replaced channel-wise
concatenation operations with simple element-wise multiplications for
each skip connection between encoder and decoder (i.e., for each
U-connection in U-Net). In our prior experiments, it reduced parameters
with negligible performance degradation.

Also, we removed the other skip connections, especially, skip
connections used in dense blocks (Choi et al., 2020; Takahashi et al.,
2018; Takahashi \& Mitsufuji, 2021, 2017). We observed that stacked
convolutional networks without dense connections followed by
Time-Distributed Fully connected layers (TDF) (Choi et al., 2020) could
perform comparably to dense blocks without TDFs. TDF, proposed in (Choi
et al., 2020), is a sequence of linear layers. It is applied to a given
input in the frequency domain to capture frequency-to-frequency
dependencies of the target source. Since a single TDF block has the
whole receptive field in terms of frequency, injecting TDF blocks into a
conventional U-Net (Ronneberger et al., 2015) improves the SDR
performance on singing voice separation even with a shallower structure.

By introducing such tricks, we found a computationally efficient and
effective model design. As a result, the proposed architecture has a
time-frequency branch and a time-domain branch, where each branch
separates stems, respectively. It blends results from two streams to
generate the final estimation. KUIELab-MDX-Net took second place on
leaderboard A and third place on leaderboard B in the Music Demixing
Challenge at ISMIR 2021. This paper also summarizes experimental results
on another benchmark, MUSDB18.

\hypertarget{background}{%
\section{Background}\label{background}}

\hypertarget{frequency-transformation-for-source-separation}{%
\subsection{Frequency Transformation for Source
Separation}\label{frequency-transformation-for-source-separation}}

Some source separation methods (Choi, 2021; Choi et al., 2020; Yin et
al., 2020) have adopted Frequency Transformation (FT) to capture
frequency-to-frequency dependencies of the target source. Both designed
their FT blocks with fully connected layers, also known as linear
layers. For example, (Choi et al., 2020) proposed Time-Distributed Fully
connected layers (TDF) to capture frequency patterns observed in
spectrograms of a singing voice. A TDF block is a sequence of two linear
layers. It is applied to a given input in the frequency domain. The
first layer downsamples the features to
\(\mathbb{R}^{\lceil F/bn \rceil}\), where we denote the number of
frequency bins in a given spectrogram feature by \(F\) and the
bottleneck factor that controls the degree of downsampling by \(bn\).

\hypertarget{tfc-tdf-u-net-v1}{%
\subsection{TFC-TDF-U-Net v1}\label{tfc-tdf-u-net-v1}}

(Choi et al., 2020) proposed the original TFC-TDF-U-Net for singing
voice separation. We call this architecture TFC-TDF-U-Net v1 for the
rest of this paper. It adopted a Time-Frequency Convolutions followed by
a TDF (TFC-TDF) block as a fundamental building block. By replacing
fully connected 2-D convolutional building blocks, conventionally used
in U-Net (Ronneberger et al., 2015) with TFC-TDF blocks, it showed a
promising performance on singing voice separation tasks of the MUSDB18
(Rafii et al., 2017b) dataset. Also, injecting TDF blocks can enhance
separation quality for the other tasks of MUSDB18, as shown in (Choi,
2021).

(Choi, 2021) presented how adding TDF blocks improves separation quality
by visualizing trained weight matrixes of single-layered TDF blocks
(they additionally trained U-Nets with single-layered TDF blocks for
weight visualization). As shown in Figure 1, each matrix is trained to
analyze timbre features uniquely observed in its instrument by capturing
harmonic patterns (i.e., \(y=\frac{\alpha}{\beta}x\)). It is also
observable that the TDF blocks still performs well on each scale.

\begin{figure}
\centering
\includegraphics{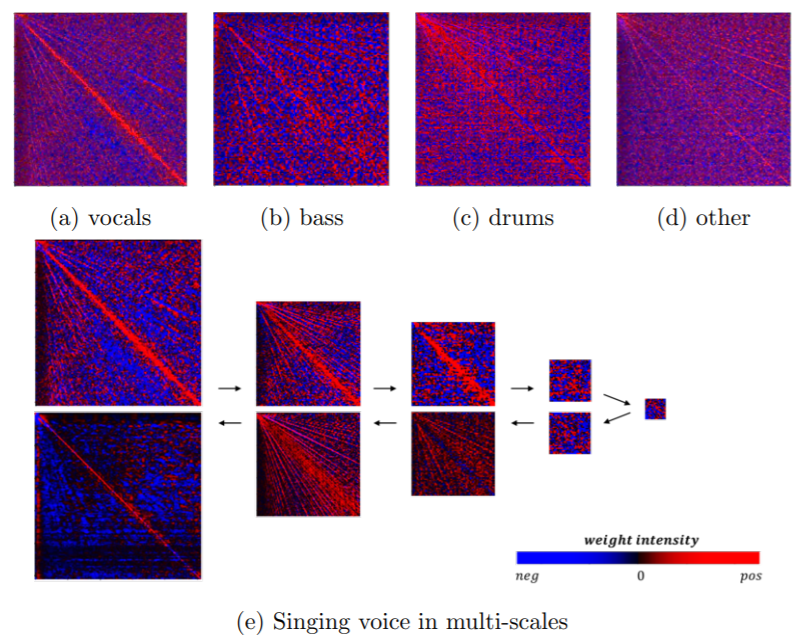}
\caption{Weight matrixes visualization of single-layered TDF blocks}
\end{figure}

We summarized TFC-TDF-U-Net v1's performance reported in (Choi, 2021) in
the experiment section.

\hypertarget{method-kuielab-mdx-net}{%
\section{Method: KUIELab-MDX-Net}\label{method-kuielab-mdx-net}}

\begin{figure}
\centering
\includegraphics{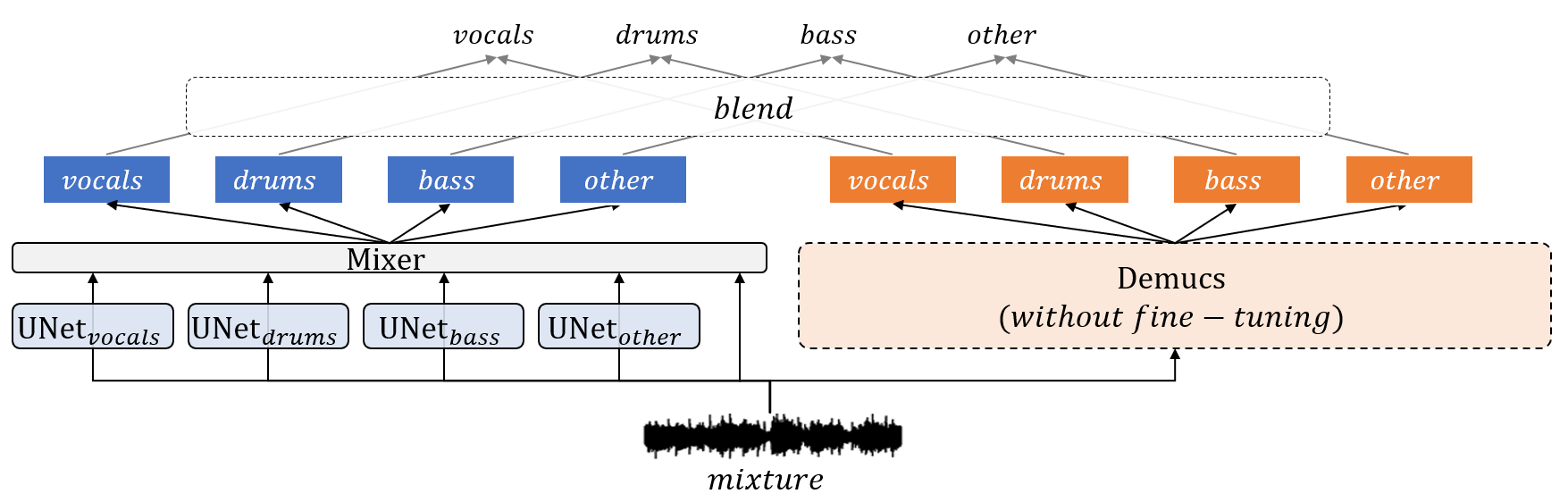}
\caption{The Overall Architecture of KUIELab-MDX-Net}
\end{figure}

Since the original TFC-TDF-U-Net v1 is computationally heavy to be
evaluated within the time limit of the MDX challenge, we could not
submit this, although its performance was promising on the MUSDB18
benchmark. To make an affordable model for the MDX challenge, we
empirically found a good balance of performance and required resources.

As in Figure 2, KUIELab-MDX-Net consists of six networks, all trained
separately. Figure 2 depicts the overall flow at inference time: the
four U-Net-based separation models (TFC-TDF-U-Net v2) first estimate
each source independently, then the \emph{Mixer} model takes these
estimated sources (+ mixture) and outputs enhanced estimated sources.
Also, we extract sources with another network based on a time-domain
approach, as shown on the right side of Figure 2. We used pretrained
Demucs (Défossez et al., 2021) without fine-tuning. Finally, it takes
the weighted average for each estimated source, also known as
\emph{blending} (Uhlich et al., 2017).

\hypertarget{tfc-tdf-u-net-v2}{%
\subsection{TFC-TDF-U-Net v2}\label{tfc-tdf-u-net-v2}}

\begin{figure}
\centering
\includegraphics{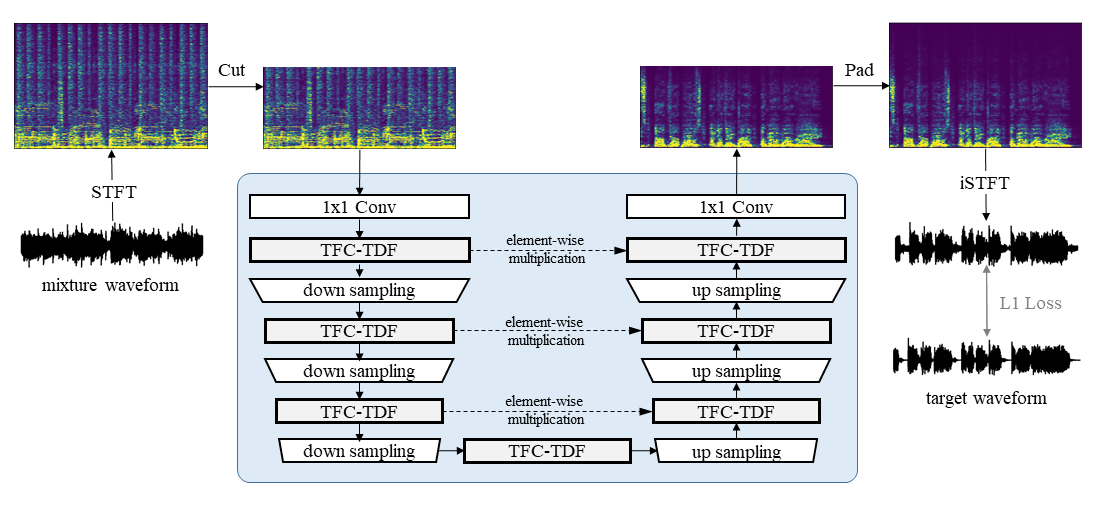}
\caption{The architecture of TFC-TDF-U-Net v2}
\end{figure}

The following changes were made to the original TFC-TDF-U-Net
architecture: - For ``U'' connections, we used multiplication instead of
concatenation. - Other than U connections, all skip connections were
removed. - In TFC-TDF-U-Net v1, the number of intermediate channels are
not changed after down/upsampling layers. For v2, they are increased
when downsampling and decreased when upsampling.

On top of these architectural changes, we also use a different loss
function (time-domain \(l_1\) loss) as well as source-specific data
preprocessing. As shown in Figure 3, high frequencies above the target
source's expected frequency range were cut off from the mixture
spectrogram. This way, we can increase \emph{n\_fft} while using the
same input spectrogram size (which we needed to constrain for the
separation time limit), and using a larger \emph{n\_fft} usually leads
to better SDR. It is also why we did not use a multi-target model (a
single model that is trained to estimate all four sources), where we
could not use source-specific frequency cutting.

\hypertarget{mixer}{%
\subsection{Mixer}\label{mixer}}

Although training one separation model for each source can benefit from
source-specific preprocessing and model configurations, these models
lack the knowledge that they are separating using the same mixture. We
thought an additional network that \emph{could} exploit this knowledge
(which we call the Mixer) could further enhance the \emph{independently}
estimated sources. For example, estimated `vocals' often have drum snare
noises left. The Mixer can learn to remove sounds from `vocals' that are
also present in the estimated `drums' or vice versa.

We only tried very shallow models (such as a single convolution layer)
for the Mixer during the MDX Challenge due to the time limit. We look
forward to trying more complex models in the future since even a single
\(1 \times 1\) convolution layer was enough to make some improvement on
total SDR (Section ``Performance on the MUSDB18 Benchmark'').

\hypertarget{experimental-results}{%
\section{Experimental Results}\label{experimental-results}}

This section describes the model configurations, STFT parameters,
training procedure, and evaluation results on the MUSDB18 benchmark. For
training, we used the MUSDB-HQ dataset with default 86/14 train and
validation splits.

\hypertarget{configurations-and-training}{%
\subsection{Configurations and
Training}\label{configurations-and-training}}

We present a comparison between configurations of TFC-TDF-U-Net v1 and
v2 as follows. This applies to all models regardless of the target
source (we did not explore different model configurations for each
source). In short, v2 is a more shallow but wider model than v1.

\begin{longtable}[]{@{}llllllll@{}}
\toprule
& \# blocks & \# convs per block & \(bn\) & \# freq bins & \# STFT
frames & hop size &\tabularnewline
\midrule
\endhead
v1 & 9 & 5 & 16 & 2048 & 128 & 1024 &\tabularnewline
v2 & 11 & 3 & 8 & 2048 & 256 & 1024 &\tabularnewline
\bottomrule
\end{longtable}

The number of intermediate channels is increased/decreased after
down/upsampling layers with a linear factor of 32. Also, as mentioned in
Section ``TFC-TDF-U-Net v2,'' we used different \emph{n\_fft} for each
source: (6144, 4096, 16384, 8192) for (vocals, drums, bass, other).

All five models (four separation models + Mixer) were optimized with
RMSProp with no momentum. We used random chunking and mixing instruments
from different songs for data augmentation (Uhlich et al., 2017). We
also used data augmentation based on pitch shift and time stretch
(Défossez et al., 2021). The overall training procedure can be
summarized into two steps:

\begin{enumerate}
\def\labelenumi{\arabic{enumi}.}
\tightlist
\item
  Train single-target separation models (TFC-TDF-U-Net v2) for each
  source.
\item
  Train the Mixer while freezing the pretrained weights of the
  separation models.
\end{enumerate}

\hypertarget{performance-on-the-musdb18-benchmark}{%
\subsection{Performance on the MUSDB18
Benchmark}\label{performance-on-the-musdb18-benchmark}}

We compare our models with current state-of-the-art models on the
MUSDB18 benchmark using the SiSEC2018 version of the SDR metric (BSS
Eval v4 framewise multi-channel SDR). We report the median SDR over all
50 songs in the MUSDB18 test set. Only models for Leaderboard A were
evaluated since our submissions for Leaderboard B uses the MUSDB18 test
set as part of the training data.

We summarize the MUSDB18 benchmark performance of KUIELab-MDX-Net. We
compare it to recent state-of-the-art models: TFC-TDF-U-Net v1 (Choi et
al., 2020), X-UMX (Sawata et al., 2021), Demucs (Défossez et al., 2021),
D3Net (Takahashi \& Mitsufuji, 2021), ResUNetDecouple+ (Kong et al.,
2021). We also include our baselines to validate our architectural
design. Even though our models were downsized for the MDX Challenge, we
can see that it gives superior performance over the state-of-the-art
models and achieves the best SDR for every instrument except `bass.'
Also, it is notable that TFC-TDF-U-Net v2 with Mixer (i.e., v2 + Mixer)
outperforms the existing methods except for `vocals' even without
blending with Demucs.

\begin{longtable}[]{@{}lllll@{}}
\toprule
& vocals & drums & bass & other\tabularnewline
\midrule
\endhead
TFC-TDF-U-Net v1 (Choi et al., 2020) & 7.98 & 6.11 & 5.94 &
5.02\tabularnewline
X-UMX (Sawata et al., 2021) & 6.61 & 6.47 & 5.43 & 4.64\tabularnewline
Demucs (Défossez et al., 2021) & 6.84 & 6.86 & 7.01 &
4.42\tabularnewline
D3Net (Takahashi \& Mitsufuji, 2021) & 7.24 & 7.01 & 5.25 &
4.53\tabularnewline
ResUNetDecouple+ (Kong et al., 2021) & 8.98 & 6.62 & 6.04 &
5.29\tabularnewline
TFC-TDF-U-Net v2 & 8.81 & 6.52 & 7.65 & 5.70\tabularnewline
v2 + Mixer & 8.91 & 7.07 & 7.33 & 5.81\tabularnewline
v2 + Demucs & 8.80 & 7.14 & \textbf{8.11} & 5.90\tabularnewline
KUIELab-MDX-Net & \textbf{9.00} & \textbf{7.33} & 7.86 &
\textbf{5.95}\tabularnewline
\bottomrule
\end{longtable}

We also compare three winning models' performance (Mitsufuji et al.,
2021) on the MUSDB18 benchmark as follows. It should be noted that we
only reported SDRs evaluated on MUSDB18 (Rafii et al., 2017a), not
MUSDB-HQ (Rafii et al., 2019).

\begin{longtable}[]{@{}lllll@{}}
\toprule
& vocals & drums & bass & other\tabularnewline
\midrule
\endhead
Hybrid Demucs (defossez) & 8.04 & \textbf{8.58} & \textbf{8.67} &
5.59\tabularnewline
KUIELab-MDX-Net (kuielab) & \textbf{9.00} & 7.33 & 7.86 &
\textbf{5.95}\tabularnewline
Danna-Sep (KazaneRyonoDanna) & 7.63 & 7.20 & 7.05 & 5.20\tabularnewline
\bottomrule
\end{longtable}

\hypertarget{acknowledgements}{%
\section{Acknowledgements}\label{acknowledgements}}

This research was supported by Basic Science Research Program through
the National Research Foundation of Korea(NRF) funded by the Ministry of
Education(NRF-2021R1A6A3A03046770). This work was also supported by the
National Research Foundation of Korea(NRF) grant funded by the Korea
government(MSIT)(No.~NRF-2020R1A2C1012624, NRF-2021R1A2C2011452).

\hypertarget{references}{%
\section*{References}\label{references}}
\addcontentsline{toc}{section}{References}

\hypertarget{refs}{}
\begin{CSLReferences}{1}{0}
\leavevmode\hypertarget{ref-choi:phd}{}%
Choi, W. (2021). \emph{Deep learning-based latent source analysis for
source-aware audio manipulation} {[}PhD thesis{]}. Korea University.

\leavevmode\hypertarget{ref-choi:2021}{}%
Choi, W., Kim, M., Chung, J., \& Jung, S. (2021). Lasaft: Latent source
attentive frequency transformation for conditioned source separation.
\emph{Proc. IEEE International Conference on Acoustics, Speech and
Signal Processing~(ICASSP)}, 171--175.
\url{https://doi.org/10.1109/ICASSP39728.2021.9413896}

\leavevmode\hypertarget{ref-choi:2020}{}%
Choi, W., Kim, M., Chung, J., Lee, D., \& Jung, S. (2020). Investigating
u-nets with various intermediate blocks for spectrogram-based singing
voice separation. \emph{Proc. International Society for Music
Information Retrieval Conference~(ISMIR)}.

\leavevmode\hypertarget{ref-defossez:2021}{}%
Défossez, A., Usunier, N., Bottou, L., \& Bach, F. (2021). \emph{Music
source separation in the waveform domain}.
\url{http://arxiv.org/abs/1911.13254}

\leavevmode\hypertarget{ref-kong:2021}{}%
Kong, Q., Cao, Y., Liu, H., Choi, K., \& Wang, Y. (2021). Decoupling
magnitude and phase estimation with deep ResUNet for music source
separation. \emph{CoRR}, \emph{abs/2109.05418}.
\url{https://arxiv.org/abs/2109.05418}

\leavevmode\hypertarget{ref-liu:2019}{}%
Liu, J.-Y., \& Yang, Y.-H. (2019). Dilated convolution with dilated GRU
for music source separation. \emph{Proceedings of the Twenty-Eighth
International Joint Conference on Artificial Intelligence, {IJCAI-19}},
4718--4724. \url{https://doi.org/10.24963/ijcai.2019/655}

\leavevmode\hypertarget{ref-mdx:2021}{}%
Mitsufuji, Y., Fabbro, G., Uhlich, S., \& Stöter, F.-R. (2021). Music
demixing challenge at ISMIR 2021. \emph{arXiv Preprint
arXiv:2108.13559}.

\leavevmode\hypertarget{ref-MUSDB18}{}%
Rafii, Z., Liutkus, A., Stöter, F.-R., Mimilakis, S. I., \& Bittner, R.
(2017a). \emph{The {MUSDB18} corpus for music separation}.
\url{https://doi.org/10.5281/zenodo.1117372}

\leavevmode\hypertarget{ref-MUSDB18HQ}{}%
Rafii, Z., Liutkus, A., Stöter, F.-R., Mimilakis, S. I., \& Bittner, R.
(2019). \emph{{MUSDB18-HQ} - an uncompressed version of MUSDB18}.
\url{https://doi.org/10.5281/zenodo.3338373}

\leavevmode\hypertarget{ref-musdb:2017}{}%
Rafii, Z., Liutkus, A., Stöter, F.-R., Mimilakis, S. I., \& Bittner, R.
(2017b). \emph{MUSDB18-a corpus for music separation}.

\leavevmode\hypertarget{ref-unet:2015}{}%
Ronneberger, O., Fischer, P., \& Brox, T. (2015). U-net: Convolutional
networks for biomedical image segmentation. \emph{International
Conference on Medical Image Computing and Computer-Assisted
Intervention}.

\leavevmode\hypertarget{ref-sawata2021all}{}%
Sawata, R., Uhlich, S., Takahashi, S., \& Mitsufuji, Y. (2021). All for
one and one for all: Improving music separation by bridging networks.
\emph{Proc. IEEE International Conference on Acoustics, Speech and
Signal Processing~(ICASSP)}, 51--55.
\url{https://doi.org/10.1109/ICASSP39728.2021.9414044}

\leavevmode\hypertarget{ref-mmdenselstm:2018}{}%
Takahashi, N., Goswami, N., \& Mitsufuji, Y. (2018). Mmdenselstm: An
efficient combination of convolutional and recurrent neural networks for
audio source separation. \emph{16th International Workshop on Acoustic
Signal Enhancement, {IWAENC} 2018, Tokyo, Japan, September 17-20, 2018},
106--110. \url{https://doi.org/10.1109/IWAENC.2018.8521383}

\leavevmode\hypertarget{ref-d3net:2021}{}%
Takahashi, N., \& Mitsufuji, Y. (2021). Densely connected multi-dilated
convolutional networks for dense prediction tasks. \emph{Proc. IEEE/CVF
Conference on Computer Vision and Pattern Recognition~(CVPR)},
993--1002.

\leavevmode\hypertarget{ref-densenet:2017}{}%
Takahashi, N., \& Mitsufuji, Y. (2017). Multi-scale multi-band densenets
for audio source separation. \emph{Workshop on Applications of Signal
Processing to Audio and Acoustics (WASPAA)}.

\leavevmode\hypertarget{ref-Uhlich17}{}%
Uhlich, S., Porcu, M., Giron, F., Enenkl, M., Kemp, T., Takahashi, N.,
\& Mitsufuji, Y. (2017). Improving music source separation based on deep
neural networks through data augmentation and network blending.
\emph{Proc. IEEE International Conference on Acoustics, Speech and
Signal Processing~(ICASSP)}, 261--265.
\url{https://doi.org/10.1109/ICASSP.2017.7952158}

\leavevmode\hypertarget{ref-phasen:2020}{}%
Yin, D., Luo, C., Xiong, Z., \& Zeng, W. (2020). PHASEN: A
phase-and-harmonics-aware speech enhancement network. \emph{Proceedings
of the AAAI Conference on Artificial Intelligence}, \emph{34},
9458--9465.

\end{CSLReferences}

\end{document}